\documentclass[prb,a4paper,final,showpacs]{revtex4}
\def\c60{\ensuremath{\mbox{C$_{60}$}}}
\def\na#1{\ensuremath{\mbox{Na$_{#1}$}}}
\usepackage{amssymb}
\usepackage{amsmath}
\begin{document}
\title{Nucleation of a sodium droplet on \boldmath \c60\unboldmath}
\author{J. Roques$^{1,2}$, F. Calvo$^1$, F. Spiegelman$^1$, and
C. Mijoule$^2$}
\affiliation{$^1$ Laboratoire de Physique Quantique, IRSAMC, Universit\'e Paul
Sabatier, 118 Route de Narbonne, F31062 Toulouse Cedex \\
$^2$ CIRIMAT, Universit\'e Paul Sabatier and Institut National Polytechnique,
118 Route de Narbonne, F31062 Toulouse Cedex}
\begin{abstract}
We investigate theoretically the progressive coating of \c60 by several
sodium atoms. Density functional calculations using a nonlocal functional
are performed for Na\c60 and \na{2}\c60 in various configurations. These data
are used to construct an empirical atomistic model in order to treat larger
sizes in a statistical and dynamical context. Fluctuating charges are
incorporated to account for charge transfer between sodium and carbon atoms.
By performing systematic global optimization in the size range $1\leq n\leq
30$, we find that \na{n}\c60 is homogeneously coated at small sizes, and
that a growing droplet is formed above $n\geq 8$. The separate effects of
single ionization and thermalization are also considered, as well as the
changes due to a strong external electric field. The present results are
discussed in the light of various experimental data.
\end{abstract}
\pacs{61.48.+c,34.70.+e,36.40.Qv}
\maketitle

\section{Introduction}
\label{sec:intro}

Bulk compounds of \c60 with alkali or alkaline earth metals can develop
particularly interesting properties such as
superconductivity.\cite{c60review,hebard,na4c60,na2c60}
At the gas phase level, the interaction between a fullerene molecule and metal
atoms has attracted a significant attention, from both
theoreticians\cite{kohanoff,ostling93,rubio,hira1,hira2,springborg,ostling97,%
aree,slanina,gong,merzynski,hamamoto} and
experimentalists.\cite{wang,martin93,zim94a,weis,zim94b,zim95,tast,frank,%
fye,palpant99,kc60,palpant01,dugourd01} 
Coverage of fullerenes with semiconductor atoms or clusters has also been
investigated.\cite{tanaka99,ohara02} For alkaline earth materials
it is now commonly accepted that many atoms homogeneously surround the \c60
molecule in spherically centred shells. For other metals, such as
lithium\cite{kohanoff,zim94b,ostling97,aree} a strong charge transfer takes
place with carbon and the resulting ionic interaction may be stronger than
the metallic bonding. As a consequence, homogeneous coating of \c60 is also
observed at low coverage. The same situation is seen
for potassium\cite{weis,martin93,ostling93,ostling97} and
rubidium\cite{weis,tast} adsorbed on \c60.
The adsorption of gold leads to a somewhat different picture, where the metal
cluster is formed next to the fullerene.\cite{palpant01}
In this case metallic bonding
dominates over the partially ionic-covalent Au--C interactions.

Here we will focus on sodium, which has been the subject of a
debate.\cite{martin93,palpant99,palpant01,dugourd01}
The results of mass spectrometry measurements, and the magic numbers inferred
from these measurements by the group of Martin,\cite{martin93} lead to the
conclusion of a regular coating of \c60. Photoelectron spectroscopy
measurements by Palpant {\em et al.}\cite{palpant99,palpant01} were
interpreted rather
similarly, these authors also providing some evidence that coating proceeds by
trimers rather than monomers. On the contrary, the measurement of electric
dipoles and polarizabilities by Dugourd {\em et al.}\cite{dugourd01} seems to
indicate
that a segregated metal droplet is formed on the surface of the fullerene.
Since the experimental conditions for the production of these clusters are
similar in the three apparatus, then mainly the interpretations differ.

Up to now, there have been essentially three theoretical investigations of
this problem, at least to our knowledge.\cite{rubio,hira2,hamamoto}
The work by Rubio {\em et al.}\cite{rubio} assumes a complete wetting of \c60
by sodium with a phenomenological two-shell jellium description. This does
not help much in elucidating the actual structure of \na{n}\c60 compounds.
The {\em ab initio}\/ calculations by Hira and Ray,\cite{hira2} performed at
the unrestricted Hartree-Fock (UHF) level, and the recent density functional
theory (DFT) calculations by Hamamoto and coworkers,\cite{hamamoto} using
the local density approximation (LDA), predict quite different results for
the physical properties and favored structures of \na{2} on \c60.

In order to address this problem, we have constructed an
empirical atomistic model, allowing extensive sampling of the configuration
space in a wide range of sizes. In turn, this enhanced sampling enables one
to achieve unconstrained global optimization as well as dynamical and finite
temperature studies. 

For this model to be physically and chemically realistic, we first carefully
performed first-principles calculations of the binding of Na and \na{2} on
\c60.
These calculations, described and discussed in the next section, are converted
into a suitable set of parameters for our model, as fully described in
Sec.~\ref{sec:model}. The main results can be separated into five categories,
which will be presented in the following order. After describing the most
stable structures found with the present model, and some of their properties,
we discuss the distinct effects of charging and thermalizing the clusters,
and the possible influence of a strong external electric field. The diffusion
dynamics
of the metal atoms over the \c60 surface is also discussed. The main results
are summarized and discussed in Sec.~\ref{sec:ccl}, in the light of the various
experimental and theoretical works available. A tentative rationalization for
the apparently contradictory observations is proposed.\cite{prl}

\section{First-principles calculations}
\label{sec:abinit}

\subsection{Methods}

Density functional theory calculations have been used to provide reference
data for the interactions between a fullerene molecule and one or two sodium
atoms. The B3LYP hybrid method proposed
by Becke\cite{becke,lyp,b3lyp}
and included in the GAUSSIAN98 package\cite{gaussian}
was used. This method is a mixture of Hartree-Fock and DFT exchange terms
associated with the gradient corrected correlation functional of Lee
{\em et al.}\cite{exlee} Calculations were made using LANL2DZ basis sets
proposed by Hay {\em et al.}\cite{hay} where the double-zeta quality
was used for the carbon all-electron calculations while effective core
potentials were used for sodium atoms to replace the ten inner electron cores.
For these atoms
the valence electrons were described with double zeta quality basis sets
proposed by the same authors.

The first part of our calculation focussed on the adsorption of a single
exohedral Na atom on the \c60 molecule. The DFT-LDA calculations
performed by Hamamoto and coworkers\cite{hamamoto} showed that the
geometry of \c60 does not change significantly during the full geometrical
optimization of \na{n}\c60 (with $1\leq n\leq 12$), compared to the case of
pure \c60. So, in the present study, the fullerene was frozen at its optimized
geometry and, for Na\c60 and \na{2}\c60, only the positions of the alkali
atoms were relaxed. We can distinguish two kinds of CC bond lengths in the
\c60 molecule. The simple C-C and the double C=C bonds used in this study were
respectively frozen at 1.45~\AA\ and 1.37~\AA, close to the experimental
values of Hedberg {\em et al.}\cite{hedberg}

\subsection{Adsorption of a single alkali atom on \boldmath\c60\unboldmath}

We label H the adsorption site located above the center of an hexagonal site,
P the site above a pentagon, BHH above the double C=C bond,
BHP above the single C-C bond, and finally T above one carbon atom (vertex).
In each case, only the radial distance $d$ from the alkali atom to the centre
of \c60
was optimized. The adsorption energies ($E_a$) of $p$ sodium atoms on \c60 were
calculated from the equation:
\begin{equation}
E_a = E(\c60)+p\times E({\rm Na}) - E(\na{p}\c60).
\label{eq:ea}
\end{equation}
Different spin multiplicities were checked in order to get the most
stable ground state.

The adsorption energy, equilibrium distance, charge transferred and dipole
moments are presented in Table~\ref{tab:nac60} for the five adsorption sites.
As can be seen, H and P are the most stable configurations with a small
preference of only 0.03~eV for the hexagon.

\def\seta{\footnotemark[1]}
\def\setb{\footnotemark[2]}
\begin{table}[htb]
\caption{Adsorption energies $E_a$, equilibrium distances Na--C,
Mulliken charges on the alkali atom, and dipole moment of Na\c60. The present
values are given in bold face.}
\label{tab:nac60}
\begin{ruledtabular}
\begin{tabular}{ccccc}
Adsorption site & $E_a$ (eV) & Na--C (\AA) & $q/e$ & $\mu$ (D) \\
\hline
H & {\bf 0.65} / 0.10\seta 2.10\setb & {\bf 2.74} / 5.08\seta 2.69\setb
& {\bf 0.87} / 1.07\setb & {\bf 14.53} / 14.63\setb \\
P & {\bf 0.62} / 0.10\seta 2.05\setb & {\bf 2.70} / 5.18\seta 2.70\setb
& {\bf 0.87} / 1.06\setb & {\bf 15.60} / 15.80\setb \\
BHH & {\bf 0.58} / 0.10\seta 1.93\setb & {\bf 2.59} / 5.00\seta 2.47\setb
& {\bf 0.87} / 1.05\setb & {\bf 17.08} / 16.54\setb \\
BHP & {\bf 0.55} / 0.10\seta 1.78\setb & {\bf 2.53} / 5.00\seta 2.46\setb
& {\bf 0.85} / 1.06\setb & {\bf 16.32} / 16.48\setb \\
T & {\bf 0.52} / 0.10\seta 1.98\setb & {\bf 2.47} / 4.50\seta 2.39\setb
& {\bf 0.85} / 1.04\setb & {\bf 16.98} / 17.10\setb \\
&&&& $16.3\pm 1.6$\footnotemark[3] \\
\end{tabular}
\end{ruledtabular}
\footnotetext[1]{Unrestricted Hartree-Fock calculations
[\protect\onlinecite{hira1}]}
\footnotetext[2]{DFT-LDA calculations \protect
[\onlinecite{hamamoto}]}
\footnotetext[3]{Experimental data \protect[\onlinecite{dugourd01,antoine}]}
\end{table}

These results are in agreement with the
calculations performed by Hamamoto {\em et al.}\cite{hamamoto} who found H
as the most stable configuration with a weak difference between H and P
of about 0.05~eV. However, the adsorption energies reported in Ref.
\onlinecite{hamamoto} are between 1.23 and 1.45 eV higher than ours.
Since in both calculations double zeta basis sets were used, this
difference is certainly due to the use of the local density approximation
in the latter work. This approximation is known to often strongly overestimate
adsorption energies.\cite{becke2} In contrast, Hira and Ray\cite{hira1}
performed unrestricted Hartree-Fock calculations of Na\c60 and predicted a
very weak adsorption energy of only 0.10~eV with no energy difference between
the five different sites. Bond sites configurations, which are probably not
true minima, show a small preference of 0.03~eV
for the C=C bond, in agreement with the findings by Hamamoto and
coworkers.\cite{hamamoto} The less stable of the 5 configurations corresponds
to the vertex site, indicating that the adsorption of a single alkali atom is
favored by a maximal coordination. 

Estimates of the barrier for diffusion of Na over \c60 can be made from the
data in Table~\ref{tab:nac60}. We find barriers of 0.07~eV between H sites
or between P sites, and 0.07 and 0.10 eV between H and P sites. Thus we
can expect a significant
mobility at room temperature, and a possible influence of temperature on the
average location of sodium atoms.

The net charge on the alkali atom, as estimated from a Mulliken population
analysis,\cite{mulliken} is also given in Table~\ref{tab:nac60}.
Our results are consistent with experimental\cite{martin93,antoine,palpant01}
and other theoretical\cite{hira1,hamamoto} studies, which predicted an
electronic transfer of about one electron from the singly occupied atomic
orbital of the alkali atom to the lowest-unoccupied molecular orbital
(LUMO) of \c60. This behavior indicates a significant
ionic bonding of the alkali
atom with the fullerene. In the unrestricted Hartree-Fock calculations
performed by Hira and Ray,\cite{hira1} no significant electronic transfer
between Na and \c60 was observed. This is correlated with a large distance
(larger than 4.5~\AA) between Na and \c60, and no actual bonding.

The optimized bond lengths between the alkali atom and the nearest carbon
atom calculated using the nonlocal functional are listed in
Table~\ref{tab:nac60}. These values show a decrease from H to
P, then BHH and BHP, and finally to T sites. Thus the Na--C distance somewhat
also reflects Na coordination.

Finally, the dipole moments are given in Table~\ref{tab:nac60} for all
five configurations. Their values range from 14.53~D for site H to 17.08~D
for site BHH. They agree well with the experimental data by Antoine {\em
et al.},\cite{antoine} who reported a dipole moment of $16.3\pm 1.6$~D.
Our results are also in good agreement with the DFT-LDA calculations of
Hamamoto and coworkers.\cite{hamamoto}

\subsection{Adsorption of two alkali atoms on \boldmath\c60\unboldmath}

We are now interested in the relative position of two alkali atoms
adsorbed on the \c60 molecule. From results of the previous part
(strong electronic transfer from alkali atom to \c60 molecule and weak
diffusion barrier on the \c60 surface), we can expect that two alkali atoms
will undergo a strong electrostatic repulsion. Two configurations
were selected, with two adjacent sites or two opposite sites, respectively.
Since the H sites are the most stable in the case of a single atom,
we have restricted the present study to hexagonal sites only. In the case
of opposite sites, only the radial distance to \c60 was optimized.
For adjacent sites, the Na--Na distance was also optimized.

\begin{table}[htb]
\caption{Adsorption energies $E_a$, equilibrium distances Na--C and Na--Na,
Mulliken charges on each alkali atom, and dipole moment of \na{2}\c60.}
\label{tab:na2c60}
\begin{ruledtabular}
\begin{tabular}{cccccc}
Adsorption site & $E_a$ (eV) & Na--C (\AA) & Na--Na (\AA) & $q/e$ & $\mu$ (D)\\
\hline
adjacent & {\bf 0.98} & {\bf 2.55} & {\bf 4.46} & {\bf 0.83} & {\bf 23.75} \\
opposite & {\bf 1.31} / 0.24\seta 4.09\setb & {\bf 2.72} / 4.48\seta &&
{\bf 0.85} / 1.05\setb & {\bf 0} \\
\end{tabular}
\end{ruledtabular}
\footnotetext[1]{Unrestricted Hartree-Fock calculations
[\protect\onlinecite{hira2}]}
\footnotetext[2]{DFT-LDA calculations \protect
[\onlinecite{hamamoto}]}
\end{table}

The adsorption energies for \na{2}\c60 [Eq.~(\ref{eq:ea})] are given in
Table~\ref{tab:na2c60}. For the two configurations, adsorption energies are
given along with the optimized Na--C and Na--Na distances, Mulliken charges on
each alkali atom, and the associated dipole moment.
The most stable configuration turns out to be the one with opposite sites,
lower by about 0.34~eV than the other one. Even though the Mulliken charges
on sodium are slightly smaller, they remain comparable to the single atom
case. This significant electronic transfer (about 0.85$e$ per atom) causes
a strong electrostatic repulsive interaction between Na atoms, which in turn
destabilizes the adjacent configuration. The present result is consistent
with the DFT-LDA calculations of Hamamoto {\em et al.},\cite{hamamoto} and
with the experimental interpretations by Martin {\em et al.}\cite{martin93}
for clusters with fewer than 7 sodium atoms.

\section{Empirical modelling}
\label{sec:model}

The calculations presented in the previous section will now be used to
parameterize an atomistic model, in order to achieve large scale sampling of
the configuration space of larger clusters and perform simulations, which
are even today beyond the possibilities of {\em ab initio}\/ approaches.

\subsection{Fluctuating charges potential}

The crucial element determining the relative extents of ionic-covalent forces
and metallic forces is charge transfer between carbon and sodium. This
transfer is expected to depend on how the metal atoms are located over \c60.
For instance, atoms closest to the \c60 surface are likely to undergo a
stronger charge transfer than the most distant ones. Therefore we cannot use
a model with fixed partial charges. A convenient answer to this problem is
provided by the fluctuating charges model, often denoted as ``fluc-q''
potential. Originating from DFT analyses,\cite{mortier} these ideas
have been simultaneously but independently applied to a variety of diatomic
molecules and some peptides by Rapp\'e and Goddard,\cite{goddard} as well as
multicharged metal
clusters by Sawada and Sugano.\cite{sawada} More recently, the group of
Berne\cite{rick} has suggested how to improve molecular dynamics simulations
with fluc-q potentials by using extended Lagrangian techniques, in a fashion
similar to the Car-Parrinello method.\cite{carpar} More rigorous derivations
from density functional theory have since been proposed,\cite{york,istkowitz}
and also a treatment within hybrid quantum mechanical/molecular mechanical
models.\cite{field} The latest developments include
higher-order electrostatic terms such as dipoles,\cite{stern} and the
treatment of polarization forces.\cite{mgo} Fluc-q potentials have been used
by Ribeiro to study the dynamics in the glass-forming liquid
Ca$_{0.4}$K$_{0.6}$(NO$_3$)$_{1.4}$\cite{cakno3} as well as the Chemla effect
in molten alkali nitrates.\cite{chemla}

Let us denote by ${\bf R}=\{{\bf r}_i,{\bf r}'_j\}$ the geometry of the
\na{n}\c60 system, the vectors ${\bf r}_i$ and ${\bf r}'_j$ representing
the 3-dimensional positions of the $n$ sodium atoms and 60 carbon atoms,
respectively. We assume that the whole cluster carries a total charge $Q$,
which may not necessarily be zero. The total potential energy $V$ of the
system is written as the sum of several contributions:
\begin{equation}
V({\bf R}) = V_{\rm Na}(\{{\bf r}_i\}) + V_{\rm C}(\{{\bf r}'_j\}) +
V_{\rm inter}({\bf R}) + V_{\rm Coul}({\bf R}).
\label{eq:vtot}
\end{equation}
$V_{\rm Na}$ in the pure metallic binding energy, which we took as an
empirical many-body potential in the second moment approximation (SMA) to
the electronic density of states in the tight-binding theory.\cite{ducastelle}
This potential involves five parameters $\xi_0$, $\varepsilon_0$, $p$, $q$,
and $r_0$:
\begin{equation}
V_{\rm Na}(\{{\bf r}_i\}) = \varepsilon_0 \sum_{i<k} e^{-p(r_{ik}/r_0-1)}
- \sum_i \left[ \xi_0^2 \sum_{k\neq i} e^{-2q(r_{ik}/r_0-1)}\right]^{1/2}
\label{eq:vna}
\end{equation}
In this equation, the sum runs over all sodium atoms, and $r_{ik}$ is the
usual distance between atoms $i$ and $k$. The initial values for the five
parameters were taken from Ref. \onlinecite{sma}.
$V_{\rm C}$ is the pure covalent
binding energy of the \c60 molecule. This term was modelled with the Tersoff
potential.\cite{tersoff} We did not attempt to change any of the original
parameters, but we added several fixed point charges to improve the
electrostatic description of this molecule. Following Schelkachelva and
Tareyeva,\cite{fixedq} 60 charges $+\delta q$ were placed on each C atom,
while 30 charges $-2\delta q$ were placed at the center of each C=C bond.
Actually the full Tersoff potential is relatively costly, and we only used it
for reoptimizing some stable structures. In the next section, the effects of
approximating \c60 as rigid will be quantitatively addressed.

The term $V_{\rm inter}$ represents the non-ionic part of the
interaction between carbon and sodium atoms. It is simply approximated as a
pairwise repulsion, involving 2 new parameters $D$ and $\beta$:
\begin{equation}
V_{\rm inter}({\bf R}) = \sum_{i\in {\rm Na}_n} \sum_{j\in {\rm C}_{60}}
De^{-\beta r_{ij}}.
\label{eq:vinter}
\end{equation}
We will assume in the following that chemical bonding between Na and C is
controlled by charge
transfer, hence we neglect other effects such as dispersion forces. In the
fluc-q model, the Coulomb electrostatic interaction is expressed as
\begin{eqnarray}
V_{\rm Coul}({\bf R}) &=& \sum_i \left[ \varepsilon_{\rm Na}q_i + \frac{1}{2}
H_{\rm Na}q_i^2 \right] + \sum_j \left[ \varepsilon_{\rm C}q'_j + \frac{1}{2}
H_{\rm C} {q'}_j^2\right] \nonumber \\
&&+ \sum_{i,j} J_{ij}q_i q'_j + \sum_{i<i'} J_{ii'} q_i q_{i'}
+ \sum_{j<j'} J_{jj'} q'_j q'_{j'},
\label{eq:vcoul}
\end{eqnarray}
where $q_i$ and $q'_j$ denote the charges carried by sodium atom $i$ and
carbon atom $j$, respectively. In Eq.~(\ref{eq:vcoul}) the labels $(i,i')$
and $(j,j')$ refer to sodium and carbon, respectively. $\varepsilon_{\rm Na}$
and $\varepsilon_{\rm C}$ are the electronegativities, $H_{\rm Na}$ and
$H_{\rm C}$ the hardnesses, and $J_{ij}$ the Coulomb interactions, which are
assumed to be explicit in the interatomic distance $r_{ij}$. In practice, we
have chosen the Ohno representation:\cite{ohno}
\begin{equation}
J_{ij}(r)=\left[ r^2 + H_{ij}^{-2} \exp(-\gamma_{ij}r^2)\right]^{-1/2}.
\label{eq:ohno}
\end{equation}
For a given geometry ${\bf R}$, the effective electronegativities of each
atom are defined by $\varepsilon=\partial V/\partial q$, which yields
\begin{equation}
\varepsilon_i = \varepsilon_{\rm Na}+H_{\rm Na}q_i + \sum_j J_{ij} q'_j
+ \sum_{i'\neq i} J_{ii'} q_{i'}
\label{eq:epsi}
\end{equation}
for sodium atoms, and a similar equation for carbon atoms. Following
Sanderson,\cite{sanderson} we now use the principle of electronegativity
equalization. This principle states that, at equilibrium, all $\varepsilon$'s
are the same in the molecule. If we denote by $\lambda$ this common value,
then the charges $\{ q_i, q'_j\}$ are such that they minimize the potential
$V_{\rm Coul}$ above, under the constraint that the total charge is
prescribed to a specific value $Q$. This minimization is equivalent to
adding the extra term $\lambda(\sum_i q_i + \sum_j q'_j -Q)$ to the potential
$V_{\rm Coul}$, and thus $\lambda$ plays the role of a Lagrange multiplier.

Finding the optimal charges amounts to solving a $(n+61)\times(n+61)$ linear
system. This system has a unique solution, as long as the hardnesses are all
strictly positive. Otherwise the quadratic form $V_{\rm Coul}$ could no longer
be minimized.

We did not attempt to refine the present model further by adding higher
order electrostatic or polarization terms and we assume that the
self-consistency of charges is sufficient to ensure a correct electrostatic
balance. Dispersion forces have been neglected as well. The interested reader
is referred to the works in Refs. \onlinecite{stern} and \onlinecite{mgo} for
more details. However, we
should mention that the fixed charges $(\delta q,-2\delta q)$, denoted as
$q_k$ in what follows, contribute to a better electrostatic description for
the whole system. The fluc-q potential can easily account for such extra
fixed charges as well as for a possible external electric field ${\bf E}$.
Both fixed charges and electric field affect the self-consistent
determination of the fluctuating charges, and the potential $V_{\rm Coul}$
should be supplemented with the term
\begin{equation}
\sum_{i,k} J_{ik} q_i q_k + \sum_{j,k} J_{j,k} q'_j q_k + \sum_i ({\bf E}
\cdot {\bf r}_i)q_i + \sum_j ({\bf E}\cdot {\bf r}_j)q'_j.
\label{eq:fixedqe}
\end{equation}

\subsection{Parameterization}

The empirical fluc-q potential described in the previous paragraph has 15
parameters, namely the five parameters of the SMA potential, the two
parameters of the Na--C repulsion, the electronegativity difference
$\varepsilon_{\rm C}-\varepsilon_{\rm Na}$, the six parameters $\gamma_{ij}$
and $H_{ij}$, which include the two hardnesses $H_{\rm Na}$ and $H_{\rm C}$,
and finally the fixed charge $\delta q$. These parameters were chosen so as
to reproduce several properties previously computed by DFT calculations, or
measured in experiments. However these data do not give much insight
into the bonding between Na and C atoms. In addition, some of the parameters,
especially those involved in the $J_{ij}(r)$ functions, can be 
in principle extracted from calculations on diatomic molecules only.
Also, while Na$_2$ and C$_2$ are well characterized experimentally and
theoretically, almost no data is available for the NaC molecule.

We have therefore achieved {\em ab initio}\/ Configuration Interaction (CI)
calculations of NaC, as well as some calculations on Na$_2$ and C$_2$ to
get the Coulomb integrals. These were taken in minimal linear combinations
of atomic orbitals basis set obtained by contracting the Gaussian-type
orbitals of atomic self-consistent field calculations in the
previous basis sets. The pointwise integrals were then fitted through the
Ohno functions, Eq.~(\ref{eq:ohno}) above. The {\em ab initio}\/ calculations
on Na$_2$ also provided estimates of the binding energy and equilibrium
distance that should be predicted by the model.

\begin{table}[htb]
\caption{Data used for fitting the empirical model with fluctuating charges.}
\label{tab:param}
\begin{ruledtabular}
\begin{tabular}{ccrr}
Property & Source & Reference & Predicted \\
\hline
$q/e$ (Na\c60) & B3LYP & 0.87 & 0.88 \\
$\mu$ (Na\c60, D) & B3LYP & 14.5 & 14.5 \\
$\Delta E$ (\na{2}\c60, eV) & B3LYP & $-0.34$ & $-0.27$ \\
$r$(\na{2}, \AA) & {\em ab initio}\seta & 3.06 & 3.34 \\
$E$(\na{2}, eV) & {\em ab initio}\seta & 0.74 & 0.58 \\
$\alpha$(\c60, \AA$^3$) & exp.\setb & 76.5 & 58.5 \\
\end{tabular}
\end{ruledtabular}
\footnotetext[1]{Data from Jeung, Ref. \protect\cite{jeung}.}
\footnotetext[2]{Data by Antoine {\em et al.} from Ref.
\protect\onlinecite{polarc60}}
\end{table}

Appropriate fits of the Coulomb integrals gave us initial guesses for the
parameters $\gamma_{ij}$ and $H_{ij}$. We then obtained the whole set of
parameters by minimizing a standard error function $\chi^2$, to reproduce
the following quantities, quoted in Table~\ref{tab:param}:
\begin{enumerate}
\item the charge transfer from sodium (0.87$e$) and the electric dipole
(14.5~D) in Na\c60, values taken from our DFT calculations;
\item the energy difference between the two \na{2}\c60 isomers with sodium
atoms on adjacent or opposite hexagonal sites ($\Delta E=-0.34$~eV from our
DFT calculations, in favor of the opposite sites location);
\item the binding energy (0.74~eV) and equilibrium distance (3.06~\AA) in
\na{2} from {\em ab initio}\/ calculations;\cite{jeung}
\item the experimental electric polarizability of \c60 (76.5~\AA$^3$, taken
from Ref.~\onlinecite{polarc60}).
\end{enumerate}

\begin{figure}[htb]
\vbox to 7cm{
\includegraphics{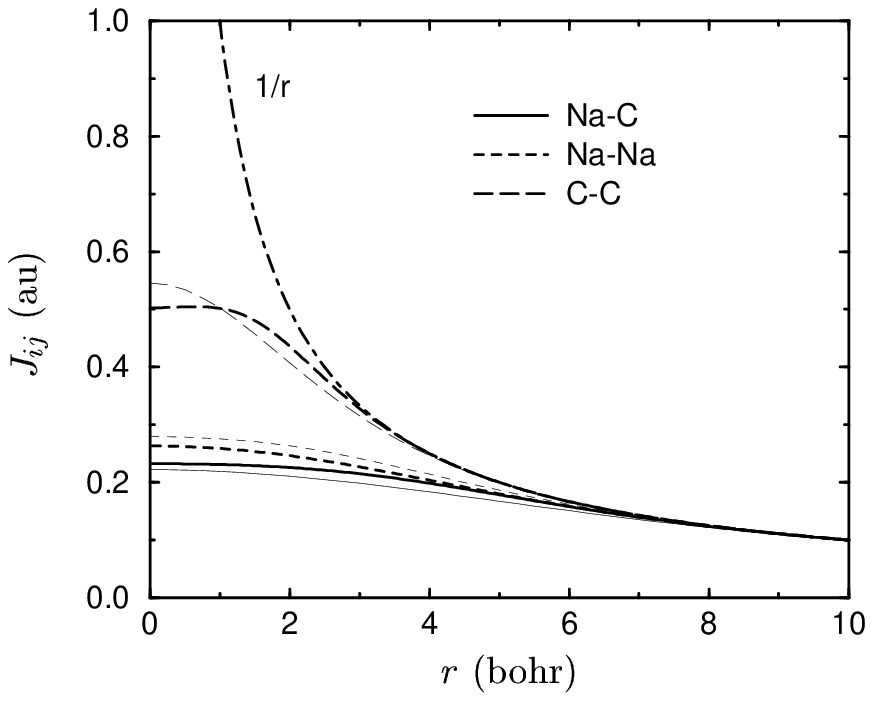}
\vfill}
\caption{Coulomb integrals for the \na{2}, C$_2$ and NaC molecules. The curves
are given for the empirical potential (thick lines) and for the {\em ab initio}
calculation (thin lines).}
\label{fig:jij}
\end{figure}

The error function $\chi^2$ was then minimized using a local Monte Carlo
search. The moves in the parameters space were not taken too large, in order to
keep the parameters close to their initial magnitudes. The final values of the
parameters of the SMA potential are $\varepsilon_0=0.01366$ eV,
$\xi_0=0.1689$ eV, $p=7.175$, $q=1.34$, and $r_0=3.29$ \AA. The parameters for
the repulsive potential between Na and C atoms are $D=313.4$ eV and
$\beta=3.423$ \AA$^{-1}$. For the Coulomb integrals, we found
$H_{\rm Na}=7.16$ eV, $H_{\rm C}=13.68$ eV, $H_{\rm NaC}=6.32$ eV,
$\gamma_{\rm Na}=0.129$ \AA$^{-2}$, $\gamma_{\rm C}=1.02$ \AA$^{-2}$, and
$\gamma_{\rm NaC}=0.15$ \AA$^{-2}$. Finally, the electronegativity difference
was taken as $\Delta \varepsilon=4.6$~eV, and the fixed charge $\delta q$ as
0.209. We represented in Fig.~\ref{fig:jij} the Coulomb interactions between
Na and C atoms. They are found to remain very close to the {\em ab initio}
calulations.

The predicted values of the quantities used in the fitting procedure are also
reported in Table~\ref{tab:param}. The general consistency is very good, even
though we note that the polarizability of \c60 is slightly underestimated in
our model. However, only the $\delta q$ parameter has some influence on this
parameter. The energy difference between the two isomers of \na{2}\c60 is
somewhat overestimated, but the crucial point here is that the isomers are
ordered similarly.

\subsection{Simulation tools}

The putative global minima, or lowest energy structures, were determined using
a variant of the Monte Carlo+minimization algorithm,\cite{bhop} also known as
basin-hopping.\cite{bhop,bh}
This optimization stage was made with a fully rigid \c60
molecule, and the random atomic displacements of the sodium atoms were of two
kinds. With a probability $p$, the selected atom is moved from its equilibrium
position by a vector $\Delta {\bf r}$ with maximum amplitude
10~\AA. With probability $1-p$, it is moved anywhere over the \c60 with the
same radial distance from the centre of mass of the fullerene. In practice, all
sodium atoms are moved before a quench is performed to find a new stable
minimum.
The value of $p$ was taken as 80\% for the smaller sizes, and decreased down
to 10\% for $n>12$. To prevent divergences in the fluc-q potential due to short
distances resulting from the random displacements, and to prevent pathological
crossing of the \c60 surface by Na atoms, we added a purely repulsive potential
between the surface of \c60 and the sodium atoms:
\begin{equation}
V_{\rm rep}(\{{\bf r}_i\}) = \sum_i \frac{\kappa}{2} (\| {\bf r}_i \| -
r_{{\rm C}_{60}})^2,
\label{eq:vrep}
\end{equation}
where we took $\kappa=880$ eV/\AA$^2$ and $r_{{\rm C}_{60}}=3.89$ \AA.
This potential was only used in concern with global optimization.

We also carried out some finite temperature molecular dynamics simulations
using the Nos\'e-Hoover chain thermostat technique.\cite{chain} In the
extended Lagrangian scheme, the charges were also thermostatted at a low
temperature $T^*=T/100$, $T$ being the vibrational temperature, to prevent
divergence of their kinetic temperature.\cite{sprik} The masses
of the thermostats were taken as $Q_i=3n k_B T/\omega^2$ (1st thermostat) and
$Q_{k>1}=Q_1/3n$ (other thermostats) for the atoms, and $Q_1^*=(n+60)k_BT^*/
\omega_0^2$ (1st thermostat) and $Q*_{k>1}=Q_1^*/(n+60)$ (other thermostats)
for the charges.\cite{chain} Appropriate values for the square frequencies
were taken as $\omega^2=10^{-6}$ and $\omega_0^2=5\times 10^{-6}$ atomic units,
respectively. The
fictitious masses of the charges were chosen equal for all charges, and taken
as $10^3$ au. The velocity Verlet algorithm was used to propagate the dynamics
with a time step $\Delta t=1$~fs, and the parallel tempering
strategy\cite{ptmc} was used to accelerate convergence when simulating the
clusters at thermal equilibrium.

\section{Results}
\label{sec:res}

\subsection{Structural properties}

\begin{figure*}[htb]
\vbox to 17cm{
\includegraphics{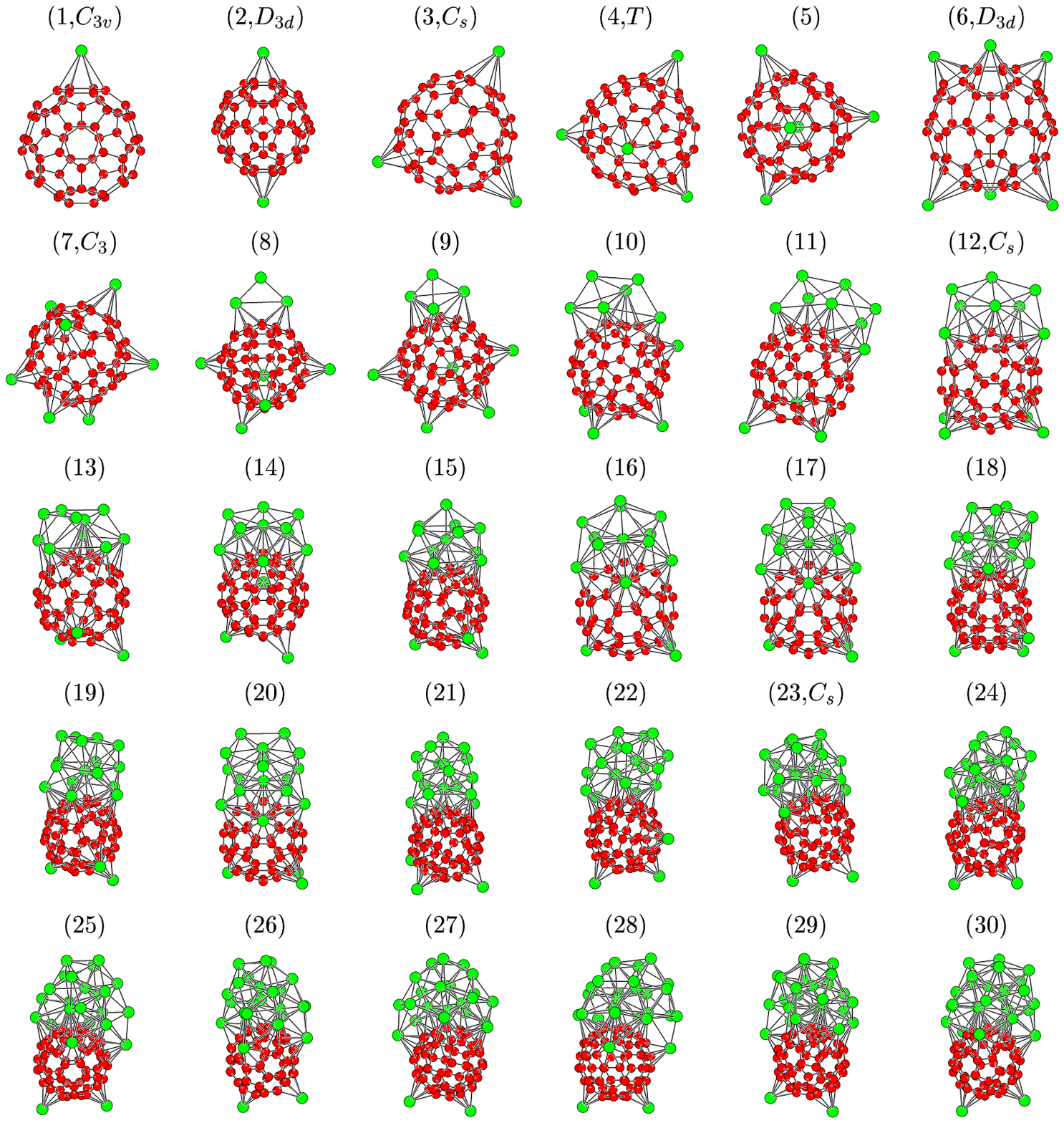}
\vfill}
\caption{Lowest-energy structures found for \na{n}\c60 in the range $1\leq n
\leq 30$ using the fluctuating charges potential.}
\label{fig:str}
\end{figure*}

For each size $n$ in the range $1\leq n\leq 30$, $10^4$ quenches were performed
during the basin-hopping searches for the global minimum. The most stable
structures found using this algorithm are reported in Fig.~\ref{fig:str}. We
have also indicated the (non-$C_1$) symmetry groups. Although this may not be
obvious from looking at Fig.~\ref{fig:str}, the \c60 molecule is also optimal
in these structures when modelled with the Tersoff potential.\cite{tersoff}
The effect of adding sodium atoms on the \c60 geometry will be quantified
later.

At low coverage, Na atoms tend to stay as far as possible from each other,
in order to minimize Coulomb repulsion induced by a significant charge
transfer. The growth roughly goes symmetric until the seventh atom is added.
There are then only few open spaces for the eighth atom to stay, and the
structure found by breaking some Na--C bonds in favor of metal Na--Na bonds
appears slightly more stable. More importantly, this droplet initially
created for $n=8$ remains at larger sizes and grows monotonically as
further atoms are deposited. Actually the droplet even captures some of the
isolated atoms during the growing process: only two atoms remain
isolated for $n=30$.

These are the main results of the present work. Our model predicts that
charge transfer is initially strong enough to inhibit metallic bonding.
Then, as more atoms are added, the competition between ionic and
metal forces finally turns at the advantage of the latter. The crossover
size for this èwetting-to-nonwetting' transition, which we estimate here to
be around $n=8$, may be of course somewhat different in more accurate chemical
descriptions. It is however worth pointing out that, even though the present
potential is not explicitely quantal, it still shows an enhanced stability
of Na trimers upon nucleation. The atom unconnected to the \c60 carries a
slightly negative charge, and the whole trimer nearly has the charge $+1$, in
agreement with the known special stability of Na$_3^+$.\cite{knight}

\begin{figure}[htb]
\vbox to 7cm{
\includegraphics{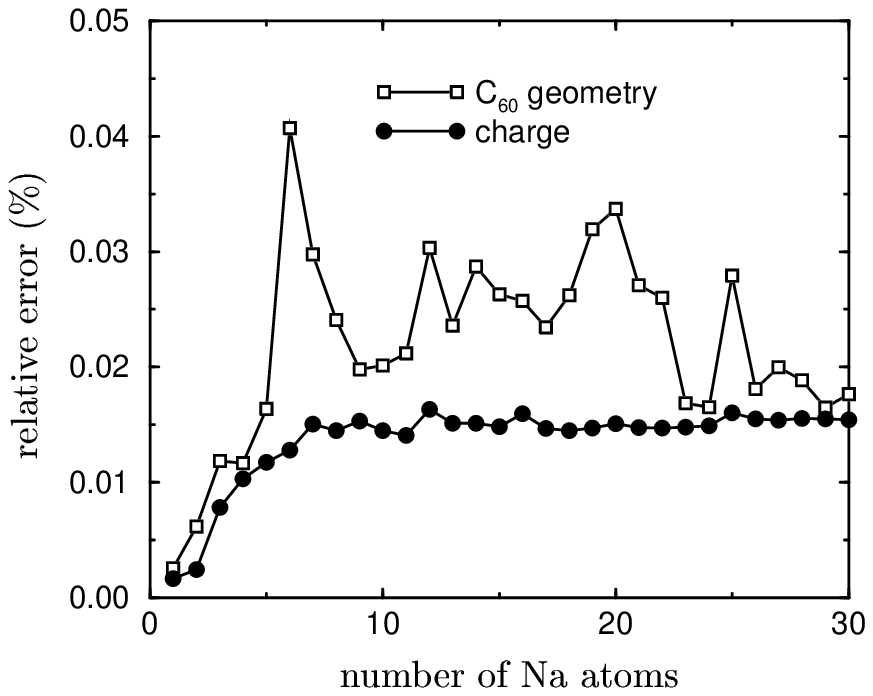}
\vfill}
\caption{Relative error for the geometric distortion (empty squares) and for
the total charge on sodium (full circles)
when considering the \c60 molecule as rigid, with
reference to the Tersoff potential.}
\label{fig:comptersoff}
\end{figure}

In Fig.~\ref{fig:comptersoff} we have plotted two indicators of the mistake
made when considering the \c60 molecule as rigid. The whole \na{n}\c60
systems were locally reoptimized using the Tersoff potential, and we calculated
the relative error $e=[A({\rm rigid})-A({\rm Tersoff})]/A({\rm Tersoff})$ for
two properties $A$: (1) the global shape is estimated from $\displaystyle
A=\sum_i \| {\bf r}_i\|^2,$
where the sum runs over all Na and C atoms, and (2) the total charge carried
by sodium atoms, $\displaystyle A=\sum_{i \in\rm Na} q_i$.

As can be seen in Fig.~\ref{fig:comptersoff}, the relative error remains
smaller than $5\times 10^{-4}$ for the shape, and
smaller than $2\times 10^{-4}$
for the charge. Of course, heating the cluster would introduce some degree of
floppyness in the fullerene. However, at the temperatures involved in the
experiment (the order of magnitude of 300~K), we do not really have to worry
about the vibrational motion of carbon atoms. The \c60 molecule itself only
melts above 2000~K.\cite{tomanek} Hence it is quite safe to use the rigid
approximation in the present work. We note, however, that other materials such
as fluorine may spontaneously induce significant deformations when they
cover \c60 in large amounts.\cite{c60f48}

\subsection{Electrostatics}

\begin{figure}[htb]
\vbox to 7cm{
\includegraphics{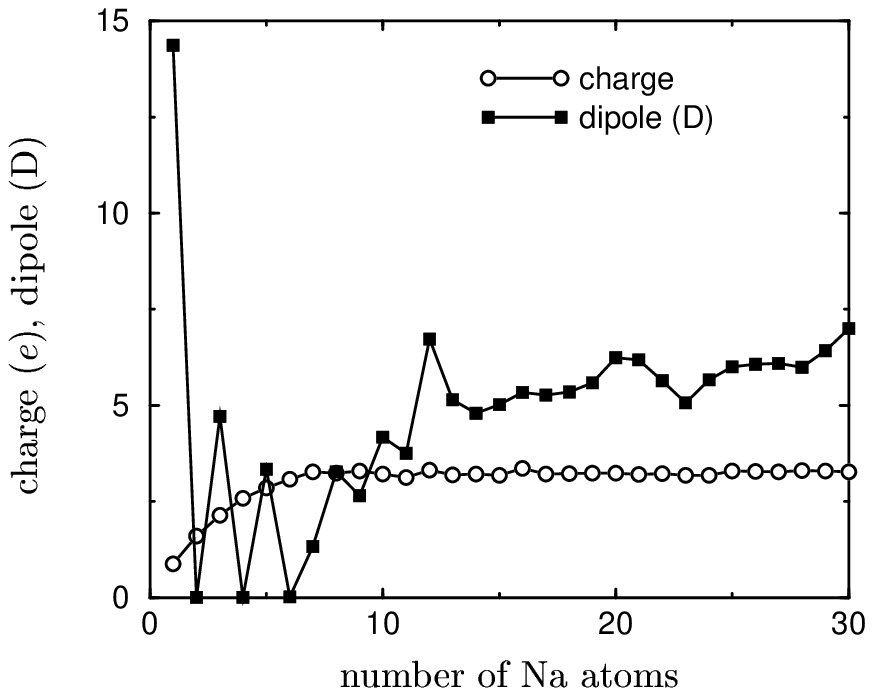}
\vfill}
\caption{Total charge (empty circles) and electric dipole moment (full
squares) versus size in \na{n}\c60 clusters.}
\label{fig:elec}
\end{figure}

The fluctuating charges potential can be straightforwardly used to compute
electrostatic properties, some of them being amenable to experimental
comparison. The total charge carried by all sodium atoms is represented in
Fig.~\ref{fig:elec} versus cluster size. As long as new hexagonal sites of
\c60 are capped by the added Na atoms, charge transfer increases steadily.
The total charge reaches a plateau at $n=8$, corresponding to the onset of
nucleation. Above this size, slightly more than 3 electrons are transferred
in the present empirical model. Single point DFT calculations performed on
some of the smaller sizes confirm this trend, but they estimate the maximal
charge transfer to be closer to 5--6 rather than 3. Such a larger value is
in agreement with the theoretical findings of Hamamoto and
coworkers.\cite{hamamoto} This may be associated with the role in the
molecular orbital calculations of the threefold degenerate LUMO orbital of
\c60, an effect which cannot be explicitely accounted for in the present
continuous electrostatic model.

In Fig.~\ref{fig:elec} we also plotted the variations of the magnitude of the
electric dipole moment $\mu$.
Two different regimes are again observed, that can
be correlated with the known structural transition. Below 8 sodium atoms, the
rather symmetric structures due to Coulomb repulsion exhibit some odd-even
alternance in the dipole. From $n=8$ and beyond, the growing droplet shows a
much more progressive increase of the dipole, except close to sizes that lose
one of the isolated atoms.

\begin{figure}[htb]
\vbox to 7cm{
\includegraphics{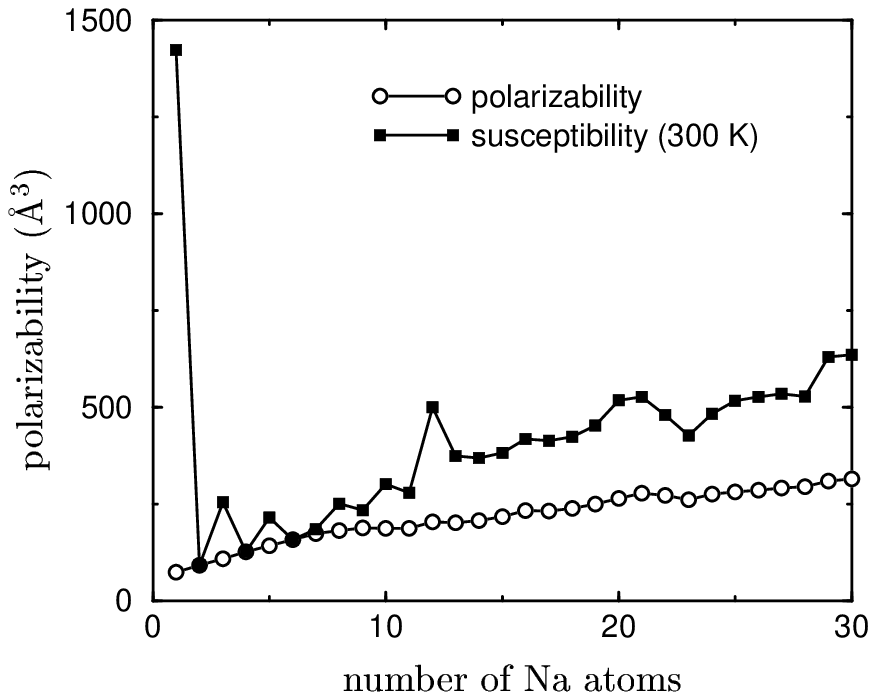}
\vfill}
\caption{Electric polarizability (empty circles) and susceptibility at 300~K
(full squares) versus size in \na{n}\c60.}
\label{fig:polar}
\end{figure}

The zero temperature electric polarizability $\alpha$
was calculated by averaging the
diagonal part of the corresponding $3\times 3$ tensor. Small electric fields
($10^{-6}$ au) were added along the three Cartesian axes, and the dipoles were
obtained after reoptimization. At finite temperature corresponding to the
experiments by Dugourd {\em et al.},\cite{dugourd01} we estimated the electric
susceptibility $\chi$ by assuming that the dipoles were rigid and statistically
oriented within the electric field. This allows us to use a simple Langevin
formula for the susceptibility, namely $\chi=\alpha + \mu^2/3k_B T$. The
variations of $\alpha$ and $\chi$(300~K) with size are represented in
Fig.~\ref{fig:polar} for the entire range $1\leq n\leq 30$.

The polarizability
shows no particular dependence with $n$ besides a steady but slow increase.
On the contrary, because $\mu$ has large variations with size, this also holds
for $\chi$. Experimental data\cite{dugourd01} are in semi-quantitative
agreement with the present results. While the general behavior observed by
Dugourd {\em et al.}\cite{dugourd01} is similar to the one for $\chi$ in
Fig.~\ref{fig:polar}, the values they measure are significantly larger. This
might be due to the neglect of probable increases in $\mu$ as temperature
increases, due to the floppyness of these clusters at 300~K.

\subsection{Charge and temperature effects}

\begin{figure}[htb]
\vbox to 7cm{
\includegraphics{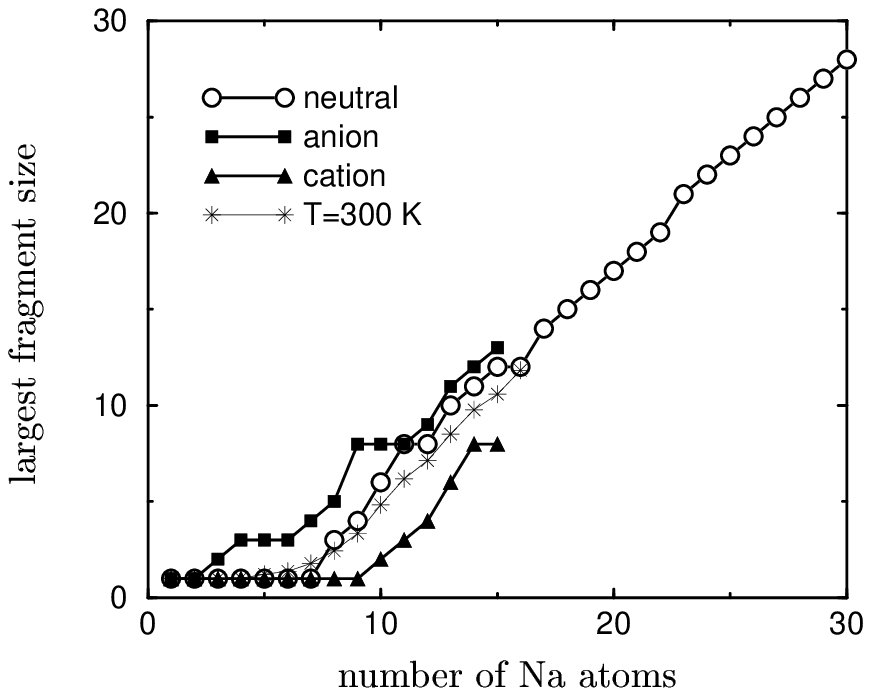}
\vfill}
\caption{Size of the largest fragment in neutral (empty circles),
anionic (full circles, $Q=-1$), cationic (full triangles, $Q=+1$)
\na{n}\c60$^Q$ clusters. The average value at 300~K calculated from
simulations is also given as stars.}
\label{fig:fragment}
\end{figure}

In experiments, clusters are ionized for subsequent size-selection. To see the
general influence of ionization on the growing process of Na clusters over
\c60, we have performed additional global optimization on anionic ($Q=-1$)
and cationic ($Q=+1$) systems. The size of the largest sodium fragment is
represented in Fig.~\ref{fig:fragment} versus the total number of sodium atoms.
Here a fragment is a set of connected atoms, where any two atoms are said to
be connected whenever they are distant by less than 8 bohr. This quantity is
suitable for detecting the onset of nucleation. Indeed, we can easily identify
in Fig.~\ref{fig:fragment} the sizes of neutral clusters where the droplet
captures one of the remaining isolated atoms. These sizes are found for
$n=13$, 18, and 22.

Cationic clusters are less stable than neutrals, as they start to nucleate a
droplet only above $n=9$. This should be attributed to the larger coulombic
repulsion the new atoms undergo as they are placed close to already present
atoms. On the contrary, charging negatively the system diminishes charge
transfer on sodium and therefore helps in creating metallic bonds. This is
precisely what we see in Fig.~\ref{fig:fragment}, where the crossover size
between coating and segregation is only 3. Obviously, the above arguments
should be further discussed in the light of quantum effects.

The finite temperature equilibrium properties have been simulated using
constant temperature, extended Lagrangian molecular dynamics as described in
the previous section. The simulations were carried out for 30 simultaneous
temperatures equidistant by $\Delta T=10$~K, and were each 10~ns long, after
2~ns were initially discarded for equilibration. To prevent the alkali atoms
from dissociating at high temperatures, we enclosed the whole system inside a
spherical container centered around the \c60.

The time averaged size of the largest fragment is also represented in
Fig.~\ref{fig:fragment}, at $T=300$~K, for $1\leq n\leq 30$. This quantity
essentially follows the 0~K behavior, but is slightly higher than 1 for
$n=6$ and 7. For $n\geq 8$, the 300~K value is slightly lower than the static
value. This suggests that some metallic bonds are weak and easily
broken for larger droplets. One must then recall that 300~K is quite a large
temperature for sodium clusters, and that dissociation events would tend to
occur on the droplet is there were no artificial container to prevent it.

\begin{figure}[htb]
\vbox to 7cm{
\includegraphics{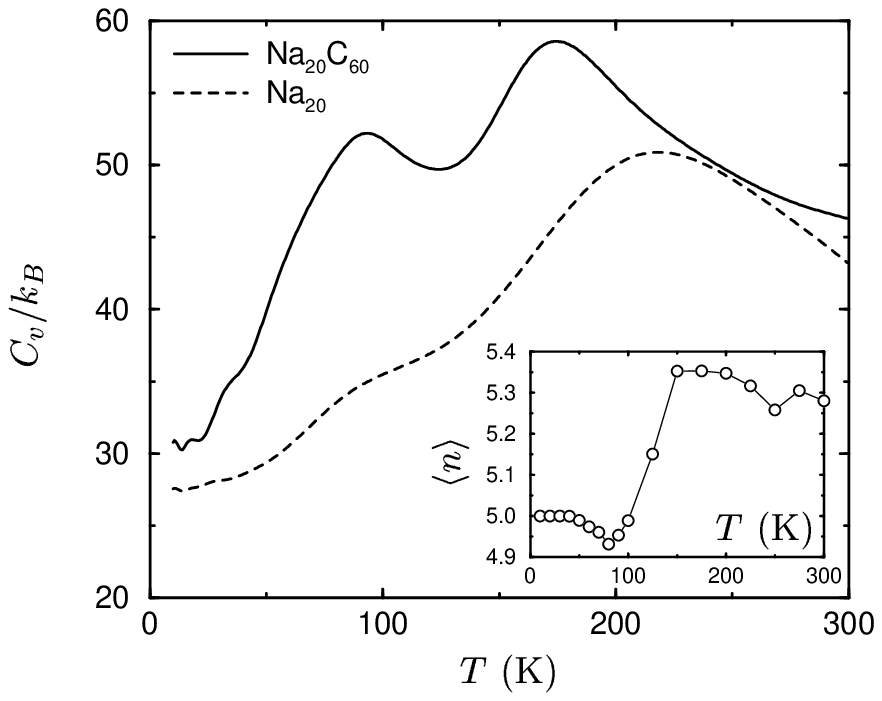}
\vfill}
\caption{Heat capacities of \na{20} in gas phase or deposited on \c60.
Inset: average fragment size $\langle n\rangle$ versus temperature in
\na{20}\c60.}
\label{fig:cv}
\end{figure}

The influence of the fullerene of the global finite-temperature behavior
of the sodium cluster was investigated by calculating the heat capacity of
\na{20}\c60, and comparing to our previous results on bare Na
clusters.\cite{prl99} The thermodynamical data was analysed using a
multiple-histogram reweighting procedure.\cite{ferrenberg}
The heat capacity curves are given in Fig.~\ref{fig:cv}, and we also
represented the variations of the (arithmetic) average fragment size $\langle
n\rangle$ with increasing temperature.

The heat capacity of \na{20}\c60 looks slightly different from that of \na{20},
with an extra shoulder bear 40~K. The main features at $\sim 100~K$ and $\sim
200~K$ in the bare cluster are enhanced in the supported cluster. 
Looking at snapshots of configurations taken from the $T=50$~K trajectory
indicates that some sodium atoms can change adsorption sites, especially
isolated ones. The peak at 100~K can be correlated with preliminar
isomerization of the \na{17} droplet, from which some atoms can dissociate
while skating over the \c60 surface. This is best evidenced on the small
decrease of $\langle n\rangle$ with $T$. On the contrary, at 180~K an inverse
process takes place, where the previously isolated atoms grow onto the droplet.
Additionally some Na$_2$ dimers may be spontaneously formed, and briefly
destroyed, during the simulation. This further increases the average fragment
size.

\subsection{Electric field}

\begin{figure}[htb]
\vbox to 11cm{
\includegraphics{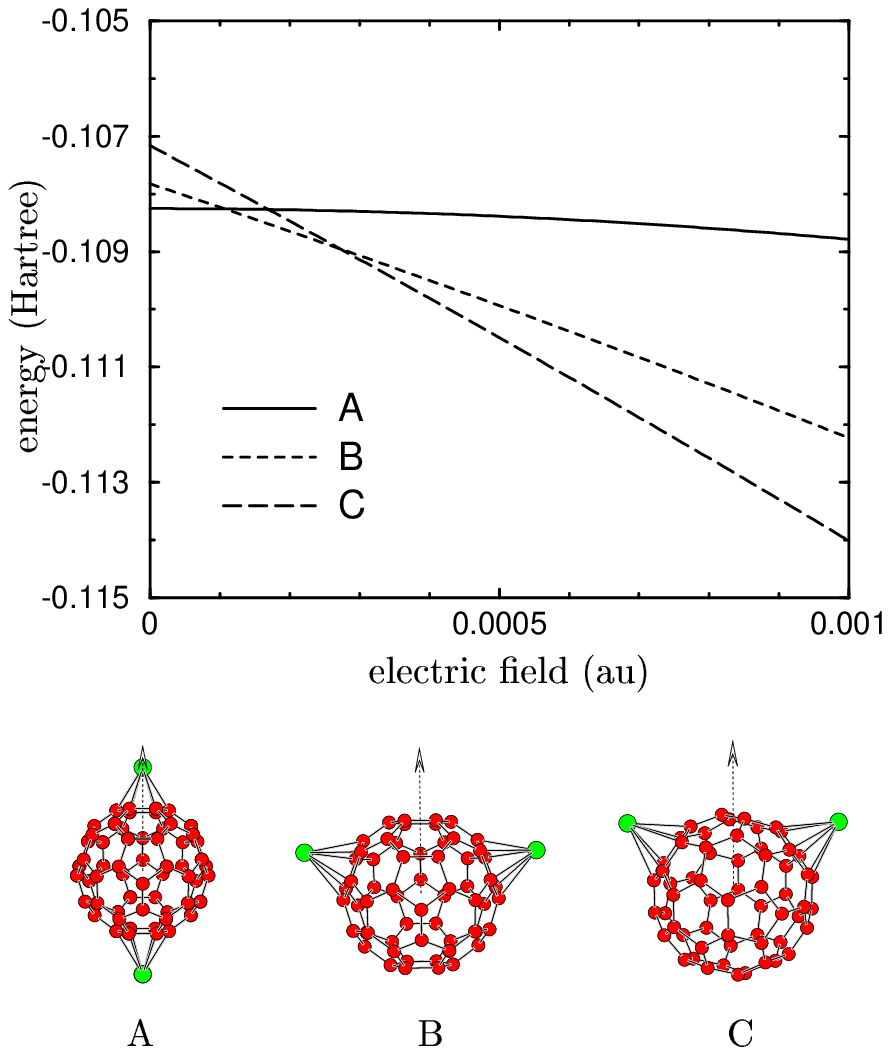}
\vfill}
\caption{Variations of the binding energy of three isomers \na{2}\c60 versus
the magnitude of electric field. The isomers (A, B, and C) are represented
on the lower panel, the arrow indicates the optimal field direction.}
\label{fig:elecisom}
\end{figure}

The experiments performed by Dugourd and coworkers\cite{dugourd01} involve the
clusters travelling through a small but very intense region of active
electric field. Actually, both the field and its gradient are strong. Because
the C and Na atoms get charged differently when put into contact in \na{n}\c60
they may react differently in presence of such a strong field. We attempted to
quantify these effects by carrying out extra global optimization for \na{2}\c60
within a field of constant magnitude $E=5\times 10^{-4}$ au, corresponding to
$2.55\times 10^8$ V~m$^{-1}$. In addition to the 4 atomic degrees of freedom
(neglecting the radial distance to the carbons), the direction of the field
must also be optimized. Using this approach we found 3 isomers to be
potentially the most stable when $E$ is waried between 0 and $10^{-3}$~au.
These isomers are represented in Fig.~\ref{fig:elecisom} along with the
variations of their energies with increasing field. In isomers B and especially
C the two alkali atoms get closer to each other. This is not surprising, since
the presence of a strong electric field causes the positively charged ions to
move in its direction. As a matter of fact, the field is optimally
perpendicular to \na{2} in both B and C, while it has little effect when fully
aligned (isomer A). Above $E\sim 10^{-4}$ au, isomer B becomes the most stable
until $E$ reached $2.5\times 10^{-4}$, above which C is the new global minimum.

The present results therefore show that a strong, external electric field
similar to the one used in experiments can indeed induce some configurational
changes in the structure of \na{n} adsorbed on \c60. Of special interest to us,
the field tends to favor metallic bonding by lowering the effects of ionic
repulsion. Hence we can expect the onset of nucleation to be earlier in the
presence of a field.

\subsection{Dynamics}

\begin{figure}[htb]
\vbox to 7cm{
\includegraphics{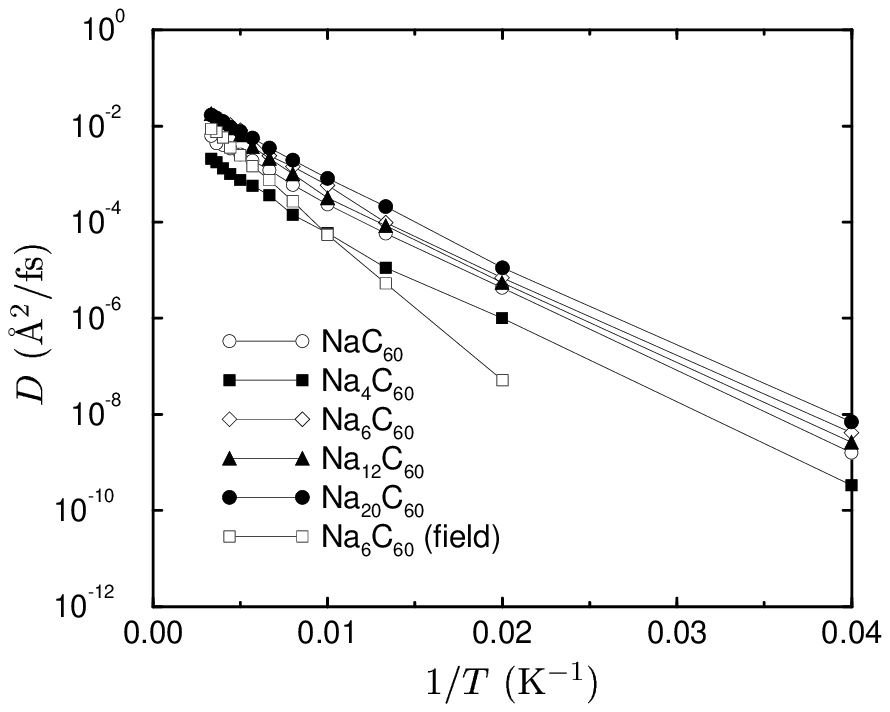}
\vfill}
\caption{Diffusion constants of various \na{n}\c60 clusters versus inverse
temperature. For \na{6}\c60, the results of simulations with non-zero external
electric field are also plotted for comparison.}
\label{fig:diff}
\end{figure}

For each temperature in the range 50~K$<T<$300~K with steps $\Delta T=50$~K,
we have performed 5000 short (10~ps long) MD trajectories from which we
computed the average atomic mean square displacement, and subsequently the
diffusion constant $D$. This was repeated for a number of sizes, and also for
a non-zero external electric field. The data for $n=1$, 4, 6, 12, and 20 are
represented as an Arrhenius plot in Fig.~\ref{fig:diff}. The general linear
variations of $\log D$ versus $1/T$ are characteristic of diffusive motion. The
activation barriers extracted from these plots are approximately $A\sim 400$~K
at zero field, and $A\sim 600$~K at $E=5\times 10^{-4}$ au. At room temperature
sodium atoms thus show a significant mobility, whatever a droplet is present
or not. Adding an electric field acts as a quench of the dynamics which could
trap the system into metastable configurations, provided that the vibrational
modes of \c60 take a part of the excess energy.

\section{Discussion and conclusion}
\label{sec:ccl}

In this work, we have performed {\em ab initio}\/ and DFT calculations on
NaC, Na\c60 and \na{2}\c60 in order to construct en empirical model for
larger clusters. This model allowed us to achieve unconstrained global
optimization as well as (thermo)dynamical investigations in a broad size
range. Our main findings are as follows:
\begin{enumerate}
\item Based on first-principles calculations, \na{2} is most stable in
dissociated form, the two sodium atoms lying on opposite hexagonal sites of
the fullerene;
\item In its initial stages, the growth of \na{n} on \c60 occurs by
minimizing Coulomb repulsion between the nearly cationic alkali atoms, thus
placing them as far as possible from each other;
\item At $n\sim 8$, the electrostatic penalty is overcome by
creating metallic bonds, and a sodium droplet is formed. The droplet grows
for larger sizes;
\item Depending on the ionic nature of the cluster, the crossover between
homogeneous coating and droplet formation varies from 3 (anion) to 10 (cation);
\item A finite temperature or a finite external electric field can enhance
metallic bonding and further decrease the crossover size;
\item At room temperature, the alkali atoms undergo a significant mobility
over the \c60 surface.
\end{enumerate}
From these results, a two-stage coating process emerges. This picture is
consistent with the observations by the Martin group\cite{martin93} and
by Palpant {\em et al.}\cite{palpant99,palpant01} who interpreted their
mass spectrometry and photoelectron spectroscopy data as the signature of
homogeneous coating at low sizes. However, in accordance with the electric
susceptibility measurements by Dugourd and coworkers,\cite{dugourd01} we
find that larger clusters are more stable when they form a droplet. Actually,
we predict that \na{n}\c60 clusters with $10<n<30$ show an intermediate
structure, which consists of a main droplet and some remaining isolated atoms.

Even though the empirical model was parameterized on {\em ab initio}
calculations and experimental data on small clusters, it is nothing more than a
model, and should thus be considered as an approximation,
especially when used for larger
clusters. Therefore, it may well be that the transition between wetted and
non-wetted forms takes place at slightly different size if we include other
effects such as polarization, or if we describe metallic bonding using a
more realistic Hamiltonian such as tight-binding. Also, the existence of
rather large sodium coverage of opposite isolated atoms, which mainly results
from electrostatic balance in avoiding too large dipole moments, might
disappear if more realistic screening due to polarization forces or quantum
effects is involved. This is to be further checked.

Nevertheless our conclusions
help in solving the experimental puzzling results of various authors.
Furthermore, it is remarkable
that very recent photodissociation and photoionization experiments carried
out by Pellarin and co-workers\cite{pellarin} on the same systems reached the
same qualitative {\em and}\/ quantitative conclusions as ours. At low coverage
($n\leq 6$) they observe
ionic-like bonding and strong charge transfer. As the number of metal atoms
increases, metallic bonds appear and become preponderant, finally controlling
cluster properties. The picture of a metallic cluster deposited on the \c60
molecule is supported by the odd-even alternation seen in the stability
pattern, where even (resp. odd) clusters evaporate mostly single atoms
(resp. dimers).

The latter effect is essentially quantum mechanical and lies beyond
the empirical potential described in the present work. The favored formation
of trimers at low coverage, suggested by Palpant and coworkers\cite{palpant01}
and further supported by the calculations of Hamamoto {\em et
al.},\cite{hamamoto} is not reproduced here. This is partially due to the
neglect of quantum effects, but also to the strong electrostatic repulsion
between the alkali atoms. At the onset of nucleation, a trimer is actually
formed, which lies perpendicular to the \c60 surface. Examination of the
partial charges shows that, while the two atoms in contact with \c60 are
significantly positively charged, the most outer atom is charged
negatively, therefore enhancing the stability of the trimer. It would be
interesting to check whether this effect is also seen in electronic structure
calculations.

Finally, we would like to stress that the present model is not limited to
simple metals adsorbed on fullerenes. Upon careful parameterization using
experimental or first-principles data, it could be suitably
modified to treat any other metal, provided that the specific metallic
interaction is taken into account. It could also be used for other materials,
which are of experimental or technological interest. For instance, the
single-crystal X-ray structure of {\c60}F$_{48}$ determined by Troyanov
and coworkers\cite{c60f48} could be reproduced using a similar bond-order
potential including fluctuating charges by Stuart {\em et al.}\cite{stuart}
More generally, it is appropriate for treating complex systems, which involve
partially covalent and metallic bonding where ionic effects and charge transfer
can prove to be a determining factor.

\section*{Acknowledgments}

The authors wish to thank Ph. Dugourd, M. Broyer, and M. Pellarin for useful
discussions. FC thanks F. Rabilloud for help with the Tersoff potential.

\appendix

\section*{Appendix: details of the configuration interaction calculation
for N\lowercase{a}C}

Some information about the distance dependence of the interaction between Na
and C in Na\c60 can be extracted from the NaC molecule. This must however be
done with care, since charge transfer is intricated with the multiplet
structure of carbon. In order to get a global picture, we have thus determined
not only the ground state but also the lowest valence excited states.

The calculation was done representing both C and Na via standard valence
semi-local
pseudopotentials of the Barthelat and Durand type\cite{barthelat}
with $[1s^2]$ and $[1s^22s^22p^6]$ cores respectively. The calculation was
carried out within the Linar Combination of Atomic Orbitals (LCAO) expansion
scheme with uncontracted gaussian type functions, namely $5s/5p/5d$ on both
centers in order to decribe the valence electrons.
%
%
A core polarization operator of the form $-1/2\alpha_c \vec{f_c}\vec{f_c}$
was added on the sodium core in order to take into account the core-valence
polarization and correlation following to the formulation of M\"uller
and Meyer\cite{muller} for the electronic contribution to the electric
field $\vec{f}$ on sodium. The Na$^+$ core polarizability was $\alpha_c
= 0.995$ a$_0^3$, and a unique stepwise cut-off radius\cite{foucrault} of
1.45 a$_0$ was used. The core-polarization contribution on carbon was
neglected due to the extremely small value of the dipole polarizability
of the C$^{4+}$ ion with respect to that of Na$^+$.

The CI treatment
was achieved using the multi-reference variational-perturbative CI
algorithm CIPSI\cite{huron} within its Quasi-Degenerate Perturbation
Theory version\cite{spiegelman} and the M\"oller-Plesset partition of
the Hamiltonian. The intermultiplet separations of the carbon atom,
calculated consistently, were found to be $\Delta(^3P-^1D)$=1.386 eV,
$\Delta(^3P-^1S)$=2.565 eV, versus 1.260 and 2.680 eV
experimentally.\cite{moore} For C$^-$ the separations were found to be 
$\Delta(^4P-^2D)$=1.656 eV, $\Delta(^4P-^2S)$=2.095 eV.

\begin{figure}[htb]
\vbox to 10cm{
\includegraphics{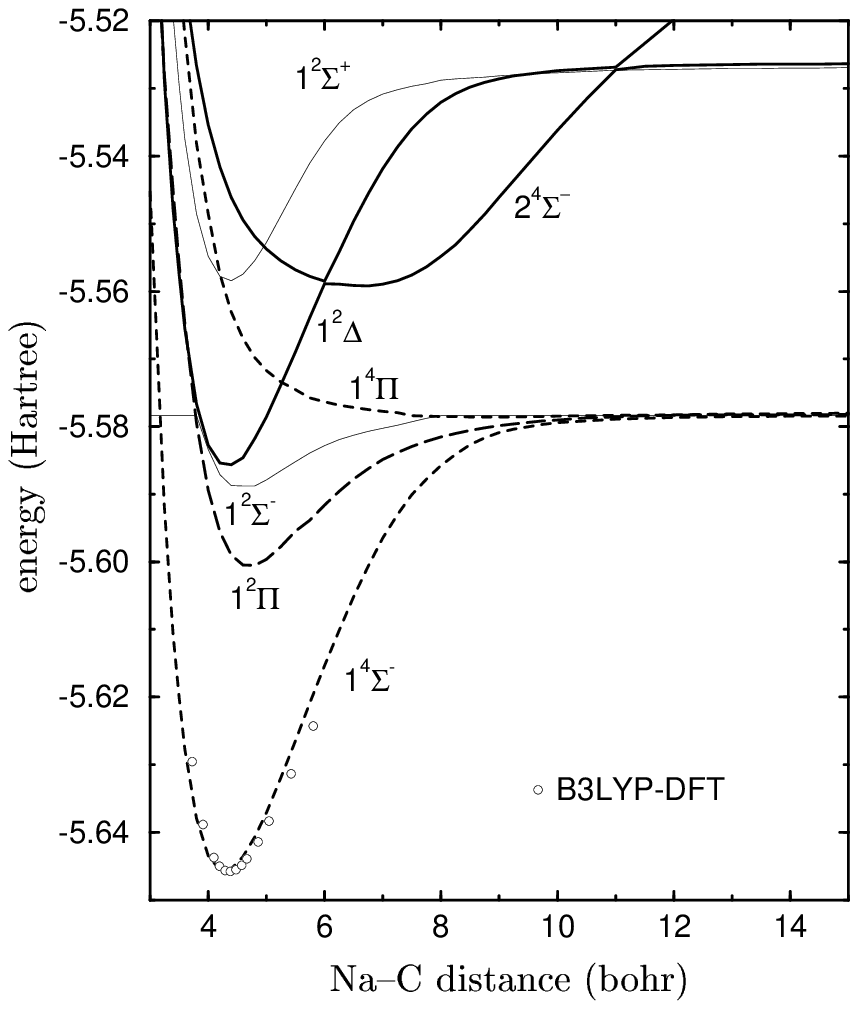}
\vfill}
\caption{Potential energy curves of the NaC molecule. All continuous curves
correspond to CI calculations. Some DFT values are also reported (dots)
for the ground state.}
\label{fig:nac}
\end{figure}

The lowest states potential energy curves are shown in Fig.~\ref{fig:nac}.
Combination of Na($3s~^2S$) with C($^3P$) generates three molecular states,
namely 1~$^4\Sigma^-$, 1~$^2\Sigma^-$ and 1~$^2\Pi$.
Combination with C($^1D$) also generates three states, 
namely 1~$^2\Delta^-$, 2~$^2 \Pi$ and 1~$^2\Sigma^+$. Finally, combination
with C($^1S$) generates a single state 2~$^2\Sigma^+$. 
In the following, the valence molecular orbitals (MO's) will be labelled
$1\sigma$, $2\sigma$, $3\sigma$, $1\pi_x$, and $1\pi_y$. Despite
hybridization, they can be qualitatively associated with the  
$2s$, $2p_z$, $3s$, $2p_x$ and $2p_y$ atomic orbitals, respectively.

The ground molecular state correlated with the C($^3P$) asymptote is a quartet
state, 1 $^4\Sigma^-$. At its equilibrium distance $R_e$=4.40 a$_0$, it
is mainly spanned by molecular configuration $1\sigma^2 2\sigma^1 1\pi_x ^1
\pi_y^1$ with all the electrons in the $2p$ shell of carbon and is therefore
expected to be strongly ionic. The DFT-B3LYP calculation of the NaC ground
state provides very similar results, see Fig.~\ref{fig:nac}. In particular,
the dipole moment is found to be 7.88~D at the equilibrium distance.
This is confirmed by the dipole moment
value $\mu=8.53$~D at the equilibrium distance and the associated
charge transfer $\delta q=0.80$. It can be clearly seen in Fig.~\ref{fig:nac}
that the ground state adiabatically undergoes at $R=7.6~a_0$ a strong
avoided crossing with a state of same symmetry 2~$^4\Sigma^-$. As a
consequence of the configuration switch, the adiabatic state 2~$^4\Sigma^-$
has a small dipole moment in the short distance range $R<6~a_0$. In the
long distance range $10~a_0< R <14~a_0$, it is correlated with
C$^-$($^4P$) + Na$^+$ and its potential curve exhibits a $-1/R$ 
behavior, consistent with a  continuously increasing dipole moment
(up to $\mu \approx -21.6$~D at $R=12~a_0$, the value
for a perfectly ionic state would be $-30.5$~D).
For still larger distances, the state interacts with another covalent
excited state. The strong coupling between the two quartet states
1~$^4\Sigma^-$ and 2~$^4\Sigma^-$ states explains the significant
dissociation energy of the ground state $D_e=1.464$ eV.

Most other states dissociating into neutral asymptotic limits show
similar (sometimes multiple) avoided crossing, mainly due to
their interaction with ionic configurations. The positions of the
avoided crossings are obviously determined by the respective positions
of the  multiplets of carbon and those of the ionic states dissociating
into C$^-$+Na. This also influences the dissociation energies of the
lowest states stabilized by the interactions. In consistency with this
picture, one can notice the avoided crossing between states
1~$^2\Sigma^-$ and 2~$^2\Sigma-$, those between the three
$^2\Pi$ states and finally that between 1~$^2\Delta$ and 2~$^2\Delta$. 
As a result, the lowest states involved in the avoided crossings are
stabilized. This leads to significant dipole moments and charge transfer
values at equilibrium for two other states correlated with the lowest
asymptote, and explains their bonding properties
($D_e=0.30$~eV, $R_e=4.6~a_0$ for state 1~$^2\Sigma^-$,
$D_e=0.60$~eV, $R_e=4.7~a_0$ for state 1~$^2\Pi$).
The same holds for state 1~$^2\Delta$
($D_e=1.61$~eV, $R_e=4.4~a_0$) dissociating into C($^1$D). 
State $1^4\Pi$ is the only exception to the previous behavior,
since it is spanned by configuration 
$1\sigma^2 2\sigma^1 1\pi_x^1 1\pi_y ^23\sigma^1$. Indeed within
a single configuration approximation, no charge transfer
can take place between the $2\sigma$ and $3\sigma$ MO's in the
quartet state because all singly occupied orbital spins are aligned.
This state can thus be described as essentially covalent with a weak dipole
moment and small charge tranfer ($\delta q=0.25$ at $R=4~a_0$).
Its potential curve is essentially non bonding and can be 
fitted via a simple exponential form, which can be used as the NaC
covalent repulsive term of the model, Eq.~(\ref{eq:vinter}).

\end{document}